\documentstyle[psfig,iopconf1]{article}
\def\be{\begin{equation}}

\def\ee{\end{equation}}

\def\dg{\mbox{$^\circ$}}		

\def\hMpc{h^{-1}{\rm Mpc}}
\def\h3Mpc{h^{-3}{\rm Mpc}^3}

\def\h3Mpcinv{h^{3}{\rm Mpc}^{-3}}

\def\spose#1{\hbox to 0pt{#1\hss}}
\def\simlt{\mathrel{\spose{\lower 3pt\hbox{$\mathchar"218$}}
     \raise 2.0pt\hbox{$\mathchar"13C$}}}
\def\simgt{\mathrel{\spose{\lower 3pt\hbox{$\mathchar"218$}}
     \raise 2.0pt\hbox{$\mathchar"13E$}}}

\input epsf
\def\plotone#1{\centering \leavevmode
\epsfxsize=\textwidth \epsfbox{#1}}
\def\plottwo#1#2{\centering \leavevmode
\epsfxsize=.45\textwidth \epsfbox{#1} \hfil
\epsfxsize=.45\textwidth \epsfbox{#2}}

\begin{document}
\title{The Sloan Digital Sky Survey and dark matter}

\author{Jon Loveday (for the SDSS collaboration)}

\affil{Astronomy \& Astrophysics Department, University of Chicago, 
5640 S Ellis Ave, Chicago, IL 60637, USA}

\beginabstract
The Sloan Digital Sky Survey (SDSS) will carry out a digital photometric
and spectroscopic survey over $\pi$ steradians in the northern Galactic cap.
An array of CCD detectors used in drift-scan mode will image the sky
in five passbands to a limiting magnitude of $r' \sim 23$.
Selected from the imaging survey, $10^6$ galaxies, $10^5$ quasars and
selected samples of stars will be observed spectroscopically.
I describe the current status of the survey, which recently saw first light,
and its prospects for constraining models for dark matter in the Universe.
\endabstract

\section{Introduction}

Systematic surveys of the local Universe can provide
some of the most important constraints on dark matter,
particularly through the measurement of the clustering of galaxies
and clusters of galaxies on large scales.
Most existing galaxy and cluster catalogues are based on photographic
plates, and there is growing concern that such surveys
might suffer from severe surface-brightness selection effects,
so that they are missing a substantial fraction of the galaxy population.
In addition, the limited volume of existing redshift surveys means
that even low-order clustering statistics, such as the galaxy two-point
correlation function, cannot reliably be measured on scales beyond
$100 \hMpc$, an order of magnitude below the scale on which COBE has
measured fluctuations in the microwave background radiation.

A collaboration has therefore been formed with the aim
of constructing a definitive map of the local universe, incorporating
digital CCD imaging over a large area in several passbands and
redshifts for around one million galaxies.
In order to complete such an ambitious project over a reasonable timescale,
it was decided to build a dedicated 2.5-metre telescope
equipped with a large CCD array imaging camera and multi-fibre spectrographs.
The Sloan Digital Sky Survey (SDSS) is a joint project of The University of
Chicago, Fermilab, the Institute for Advanced Study, the Japan
Participation Group, The Johns Hopkins University, Princeton University,
the United States Naval Observatory, and the University of Washington.
Apache Point Observatory, site of the SDSS, is operated by the Astrophysical
Research Consortium.   Funding for the project has been provided by
the Alfred P. Sloan Foundation, the SDSS member institutions, the
National Science Foundation and the U.S. Department of Energy.

\section{Survey Overview}

\begin{figure}
\plotone{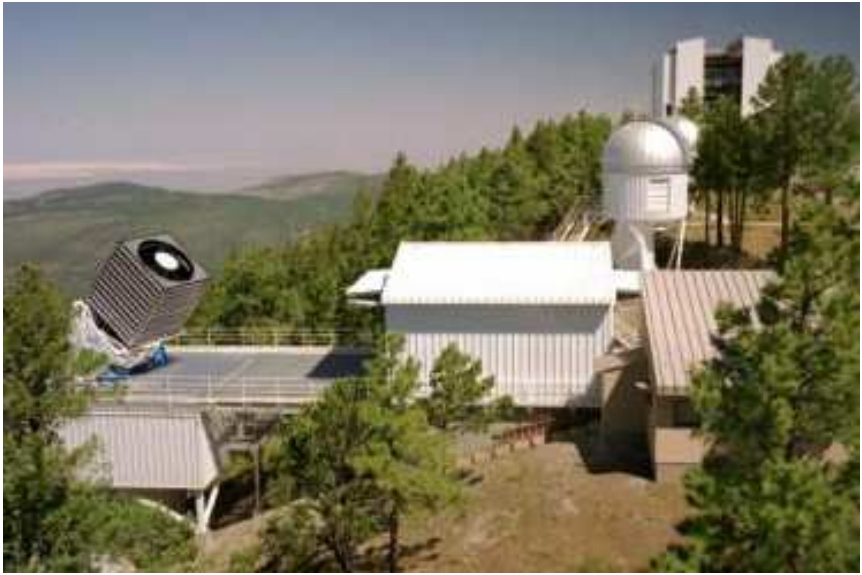}
\caption[]{View of Apache Point Observatory, with the SDSS 2.5m telescope
	to the left, it's roll-off enclosure in the centre and with the
	monitor telescope dome behind.
	The ARC 3.5m telescope is seen
	at the top right.}
\label{fig:site}
\end{figure}

\begin{figure}
\vspace{-1cm}
{\centering \leavevmode
\psfig{figure=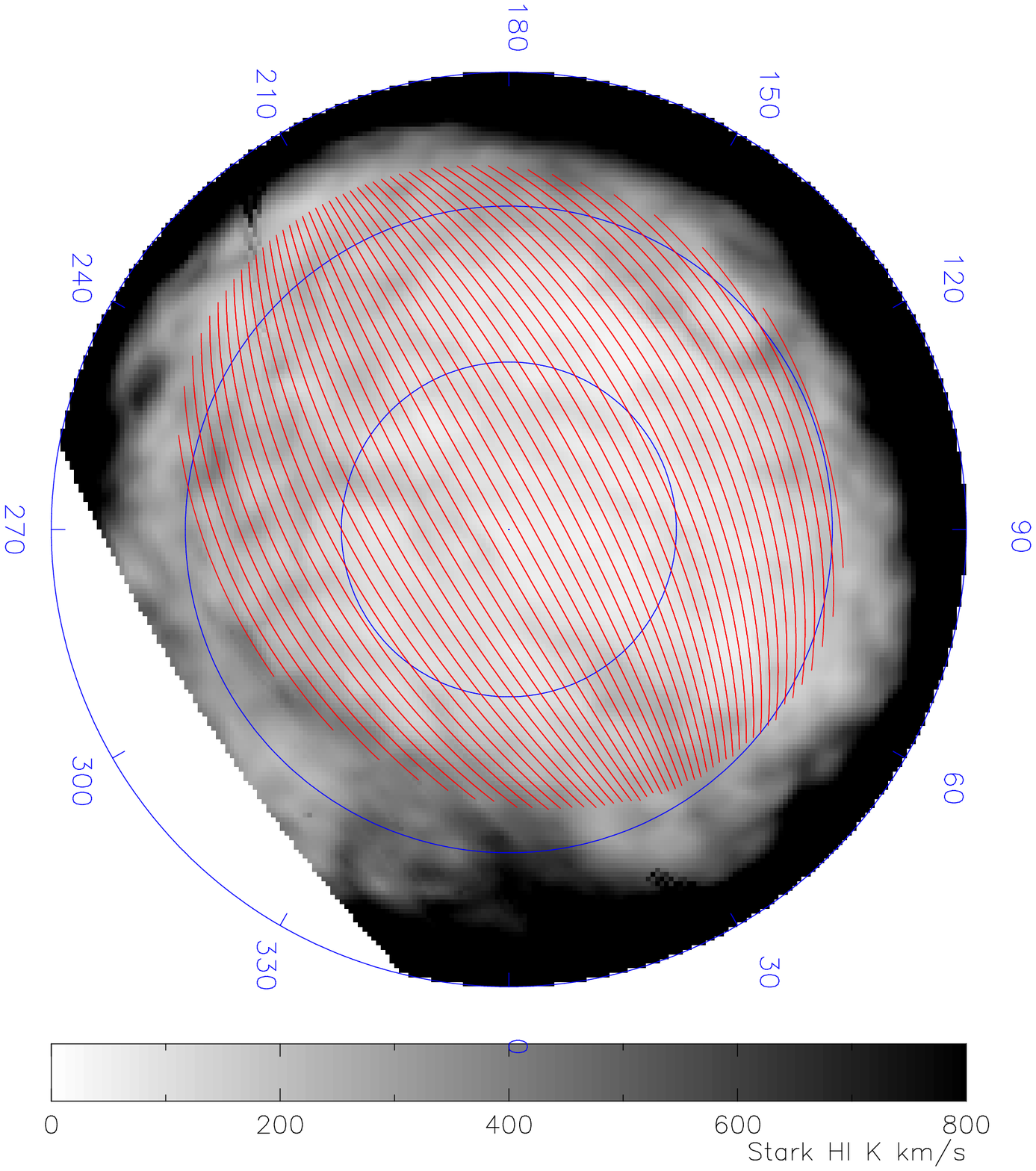,width=.49\textwidth,angle=180}
\psfig{figure=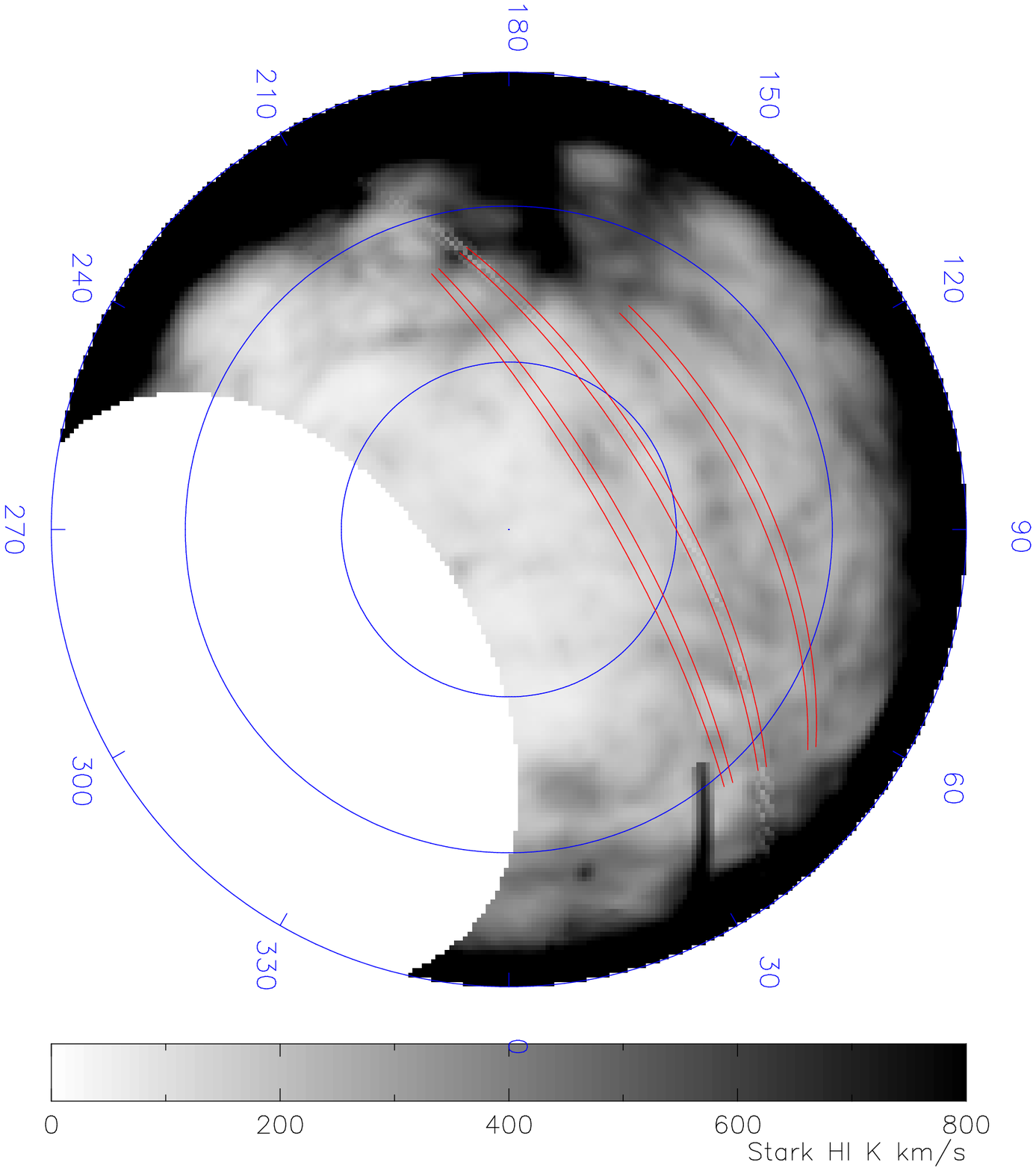,width=.49\textwidth,angle=180}
}
\vspace{-8mm}
\caption[]{Location of SDSS imaging scans in the northern (left) and southern 
	(right) galactic hemispheres.
	The concentric circles show galactic latitudes of $b = 0$, $30$ and 
	$60\dg$.
	The grey scale map shows Stark HI column density in units of
	$10^{20}$ cm$^{-2}$.
	The dark lines show the survey scan-lines, all of which follow 
	great circles.
	We observe a contiguous area of $\pi$ sr in the north, and
	three separated stripes in the south.
	Note that the northern survey is tilted with respect to the
	$b = +30\dg$ contour to avoid regions of high HI column density.}
\label{fig:stripes}
\end{figure}

Apache Point Observatory sits at 2800 metres elevation in the Sacramento
Mountains of New Mexico (Fig.~\ref{fig:site}).
The survey hardware comprises the main 2.5-metre telescope, equipped with
CCD imaging camera and multi-fibre spectrographs, a 0.5-metre monitor
telescope and a $10\mu$ all-sky camera.
On the best nights (new moon, photometric, sub-arcsecond seeing)
the 2.5-metre telescope will operate in imaging mode, drift scanning the
sky at sidereal rate, and obtaining nearly simultaneous images
in the five survey bands $u'$, $g'$, $r'$, $i'$ and $z'$ \cite{FUK}.
On sub-optimal nights, which will comprise the bulk of observing time,
the imaging camera will be replaced with a spectroscopic fibre plug-plate
cartridge.
It is planned that imaging data will be reduced and calibrated, spectroscopic
targets selected, and plates drilled within the one-month lunar cycle,
so that we will be obtaining spectra of objects that were imaged the
previous month.
We will spend most of the time observing within a contiguous $\pi$
steradian area in the north Galactic cap (NGC).
For those times when the NGC is unavailable, about one third of the time,
we will instead observe
three stripes in the southern sky, nominally centred at RA $\alpha = 5\dg$, and
with central declinations of $\delta = +15\dg$, $0\dg$ and $-10\dg$.
The southern equatorial stripe will be observed multiple times.
The location of survey scans is shown in Figure~\ref{fig:stripes}.

In the remainder of this section I discuss the various components
of the survey in more detail.

{\bf 2.5-metre telescope.}
The main 2.5-metre telescope is of modified Ritchey-Chr\'etien design \cite{wmgk}
with a $3\dg$ field of view, and is optimised for both a wide-area
imaging survey and a multi-fibre spectroscopic survey of galaxies to
$r' \sim 18$.
One of the most unusual aspects of the telescope is its enclosure.
Rather than sitting inside a dome, as is the case with conventional
optical telescopes, the enclosure is a rectangular frame structure
mounted on wheels, which is rolled away from the telescope in order
to take observations.
By completely removing the enclosure from the telescope, we can avoid
the substantial degradation to image quality due to dome seeing.
The telescope is situated on a pier overlooking a steep dropoff
so that the prevailing wind will flow smoothly over the telescope
in a laminar flow, which will also help to ensure good image quality.
A wind baffle closely surrounds the telescope, and is independently
mounted and driven.
This baffle serves to protect the telscope from stray light as
well as from wind buffeting.


{\bf Imaging Camera.}
In order to image a large area of sky in a short time, we have built
an imaging camera \cite{JEG98}
that contains $30 \times 2048^2$ photometric CCDs, arranged in six columns.
Each column occupies its own dewar and contains one chip in each of
the five filters.  Pixel size is $0.4''$.
The camera operates in drift-scan mode: a star or galaxy image drifts down
the column through the five filters, spending about 55 seconds in each.
This mode of observing has two significant advantages over conventional
tracking mode.
1) It makes extremely efficient use of observing time, since there is no
overhead between exposures: on a good night we can open the shutter,
drift-scan for eight hours and then close the shutter.
2) Since each image traverses a whole column of pixels on each CCD,
flat-fielding becomes a one-dimensional problem, and so can be
done to lower surface-brightness limits than with tracking mode images.
This, along with the high quantum efficiency of modern CCDs, will enable us
to detect galaxies of much lower surface brightness than can wide-field
photographic surveys.
There is a gap between each column of CCDs, but this gap is slightly smaller
than the width of the light-sensitive area of the CCDs, and so having
observed six narrow strips of sky one night, we can observe an interleaving
set of strips a later night, and thus build up a large contiguous area of sky.
The northern survey comprises 45 pairs of interleaving great circle scans, 
and so
imaging observations for the north will require the equivalent of 90
full photometric nights.
The camera also includes 24 smaller CCDs arranged above and below the 
photometric columns.  These extra CCDs, equipped with neutral density filters,
are used for astrometric calibration,
as most astrometric standards will saturate on the photometric CCDs.
Thus the photometric data can be tied to the fundamental astrometric
reference frames defined by bright stars.

{\bf Spectrographs.}
The 2.5-metre telescope is also equipped with a pair of fibre-fed,
dual-beam spectrographs, each with two cameras, two gratings and two $2048^2$
CCD detectors.
The blue channel will cover the wavelength range 3900--6100 \AA\ and
the red channel 5900--9100 \AA\ and both will have a spectral resolving
power $\lambda/\Delta \lambda \approx 1800$.
The fibres are $3''$ in diameter and the two spectrographs each hold 320 fibres.
Rather than employing robotic fibre positioners to place the fibres
in the focal plane, we will instead drill aluminium plates for each
spectroscopic field and plug the fibres by hand.
We plan on spectroscopic exposure times of 45 minutes and allow 15 minutes
overhead per fibre plate.
On a clear winter's night we can thus obtain 9 plates $\times$ 640 fibres
$= 5760$ spectra.
In order to allow such rapid turnaround between exposures we will assemble
9 fibre cartridges, so that each plate can be plugged 
with fibres during the day.
It will not be necessary to plug each fibre in any particular hole,
as a fibre mapping system has been built which will automatically map
fibre number onto position in the focal plane after the plate has been plugged.
This should considerably ease the job of the fibre pluggers, and we expect
that it will take well under one hour to plug each plate.

{\bf Monitor telescope.}
In order to check that observing conditions are photometric,
and to tie imaging observations to a set of primary photometric standards,
we are also employing a monitor telescope.
While the 2.5-metre telescope is drift-scanning the sky,
the 0.5-metre monitor telescope, situated close by,
will interleave observations of standard stars with calibration patches
in the area of sky being scanned.
Operation of this telescope will be completely automated, and each hour will
observe three calibration patches plus standard stars in all five colours.

{\bf $10\mu$ all-sky camera.}
As an additional check on observing conditions, a $10\mu$ infrared
camera will survey the entire sky every 10 minutes or so.
Light cirrus, which is very hard to see on a dark night, is bright
at $10\mu$, and so this camera will provide rapid warning of
increasing cloud cover, thus enabling us to switch to spectroscopic
observing rather than taking non-photometric imaging data.


{\bf Data-reduction pipelines.}
The last, but by no means least, component of the survey is a suite
of automated data-reduction pipelines
which will read DLT tapes mailed
to Fermilab from the mountain and yield reduced and calibrated data
with the minimum of human intervention.
Such software is very necessary when one considers that the imaging camera
will produce data at the rate of around 16 Gbytes per hour!

Pipelines exist to reduce each source of data from the mountain
(photometric frames and ``postage stamps'', astrometric frames,
monitor telescope frames and 2-D spectra) as well as to perform
tasks such as spectroscopic target selection and ``adaptive tiling''
to work out the optimal placing of spectroscopic field centres to maximize
the number of spectra obtained.
The pipelines are integrated into a purpose-written environment
known as Dervish
and the reduced data will be written into an object-oriented database.

\section{Data Products}

The raw imaging data in five colours for the $\pi$ steradians of the 
northern sky will occupy about 12 Tbytes uncompressed, 
but it is expected that very
few projects will need to access the raw data, which will probably
be stored only on magnetic tape.
Since most of the sky is blank to $r' \sim 23$, all detected images
can be stored, using suitable compression, in around 80 Gbytes,
and it is expected that these ``atlas images'' can be kept on spinning disc.
The photometric reduction pipeline will meaure a set of parameters
for each image, and it is estimated that the parameter lists for all objects
will occupy $\sim 15$ Gbyte.
The parameter lists for the spectroscopic sample will probably fit
into 400Mb, and the spectra themselves will occupy $\sim 25$ Gb.
Final access to the data will be through an astronomer-friendly interface, 
which will answer such queries as ``Return all galaxies
with $(g' - r') < 0.5$ and within 30 arcminutes of this quasar'', etc.

\subsection{Spectroscopic Samples}

The spectroscopic sample is divided into several classes.
In a survey of this magnitude, it is important that the selection 
criteria for each class remain fixed throughout the duration of the survey.
Therefore, we will spend a considerable time (maybe one year), obtaining
test data with the survey instruments and refining the spectroscopic selection
criteria in light of our test data.
Then, once the survey proper has commenced, these criteria will be
``frozen in'' for the duration of the survey.
The numbers discussed below are therefore only preliminary, and we expect
them to change slightly during the test year.

The {\bf main galaxy sample} will consist of $\sim 900,000$ galaxies selected
by Petrosian magnitude in the $r'$ band, $r' \simlt 18$.
Simulations have shown that the Petrosian magnitude,
which is based on an aperture defined by the ratio of light within an annulus 
to total light inside that radius, provides probably the least biased
and most stable estimate of total magnitude.
There will also be a surface-brightness limit, so that we do not
waste fibres on galaxies of too low surface brightness to give a reasonable
spectrum.
This galaxy sample will have a median redshift $\langle z \rangle \approx 0.1$.

We plan to observe an additional $\sim 100,000$
{\bf luminous red galaxies} to $r' \simlt 19.5$.
Given photometry in the five survey bands, redshifts can be estimated
for the reddest galaxies to $\Delta z \approx 0.02$ or better \cite{c95},
and so one can also predict their luminosity quite accurately.
Selecting luminous red galaxies, many of which will be cD galaxies in cluster 
cores, provides a valuable supplement to the main
galaxy sample since 1) they will have distinctive spectral features,
allowing a redshift to be measured up to 1.5 mag fainter than the main sample,
and 2) they will form an approximately volume-limited sample with a
median redshift $\langle z \rangle \approx 0.5$.
They will thus provide an extremely powerful sample for studying
clustering on the largest scales and the evolution of galaxies.

{\bf Quasar} candidates will be selected by making cuts
in multi-colour space and from the FIRST radio catalogue \cite{bwh95},
with the aim of observing $\sim 100,000$ quasars.
This sample will be orders of magnitude larger than any existing quasar
catalogues, and will be invaluable for quasar luminosity function, evolution
and clustering studies
as well as providing sources for followup absorption-line observations.

In addition to the above three classes of spectroscopic sources, which are
designed to provide {\em statistically complete} samples, we will also 
obtain spectra
for many thousands of {\bf stars} and for various {\bf serendipitous}
objects.
The latter class will include objects of unusual colour or morphology
which do not fit into the earlier classes, plus unusual objects found
by other surveys and in other wavebands.

\section{Current Status}

\begin{figure}
\plotone{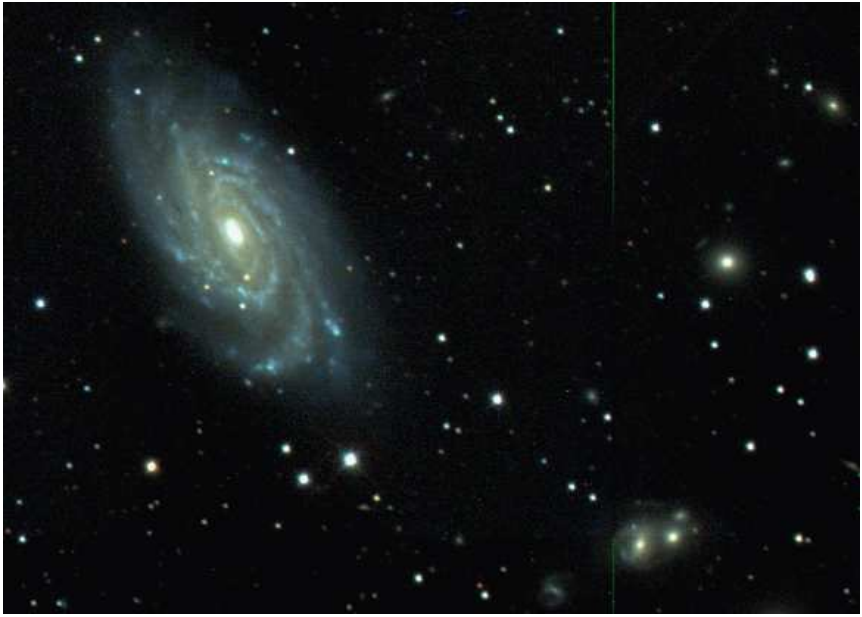}
\caption[]{A small section of the first-light image obtained on the
           night of May 27-28, 1998 which includes the galaxy NGC 6070.} 
\label{fig:first_light}
\end{figure}

The 2.5m telescope and imaging camera are complete and in place.
First light with the imaging camera was obtained on 9 May 1998
during bright time and without the baffles; subsequently three imaging runs
have been made during dark time and with the baffles in place.
We obtained sub-arcsecond images on our most recent run (September 1998).
A small sample of SDSS imaging data is shown in Figure~\ref{fig:first_light}.
This and other colour images are available from the survey website:
{\tt http://www.sdss.org/}.

The spectrographs are both at the site and all optics have been
completed and coated.  One spectrograph has been fully assembled and
some test spectra taken.  A fully assembled fibre cartridge is 
ready and all the others are ready for assembly.  
The full complement of over 6,000 science fibres needed for
the survey have been accepted and tested with a mean throughput of 92.0\%.
Test plug plates have been drilled with all positions well within
tolerances.  The various pieces of equipment for storing, handling, and
transporting the cartridges are all in place.
Once the telescope control system is in operation later this year, 
we will be able to take spectra on the sky.

All of the data reduction-pipelines are written, with 
ongoing work on minor bug-fixes, speed-ups and integration of the entire
data processing system.
The imaging/photometric reduction pipelines are being exercised with
the data taken this May and June by the survey imaging camera.
Tests are being carried out on the spectroscopic reduction pipeline using
simulated data as well as daylight test spectra.

The intent of this project is to make the survey data available to the
astronomical community in a timely fashion. We currently plan to
distribute the data from the first two years of the survey no later
than two years after it is taken, and the full survey no later than
two years after it is finished.
The first partial release may or may
not be in its final form, depending on our ability to calibrate it
fully at the time of the release. The same remarks apply to the
release of the full data set, but we expect the calibration effort to
be finished before that release.

\section{Prospects for constraining dark matter}

Since the main topic of this meeting is dark matter, I will highlight
two of the areas in which the SDSS will provide valuable data for
constraining dark matter.

\subsection{Measurement of the Fluctuation Spectrum}

\begin{figure}
\plottwo{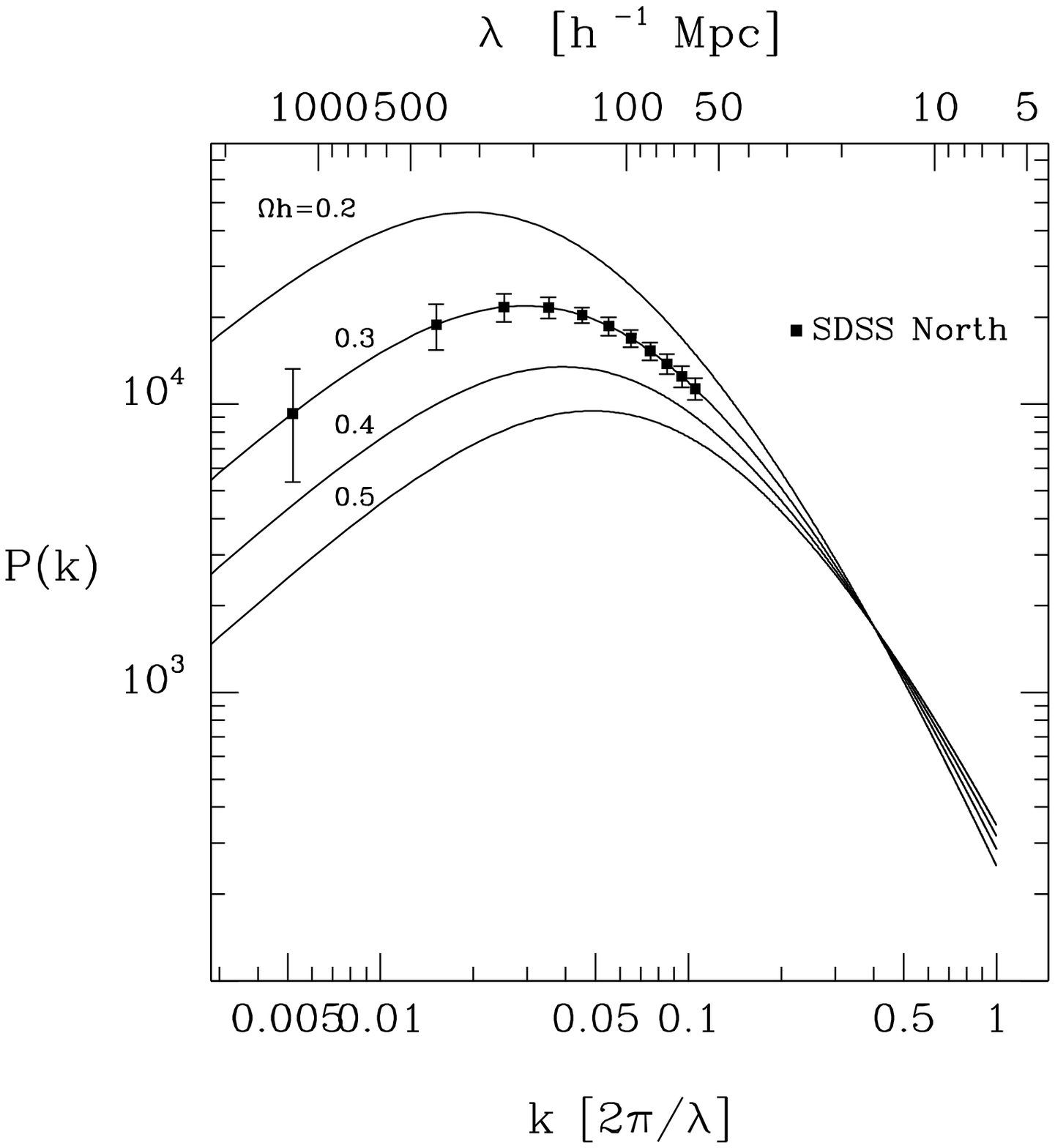}{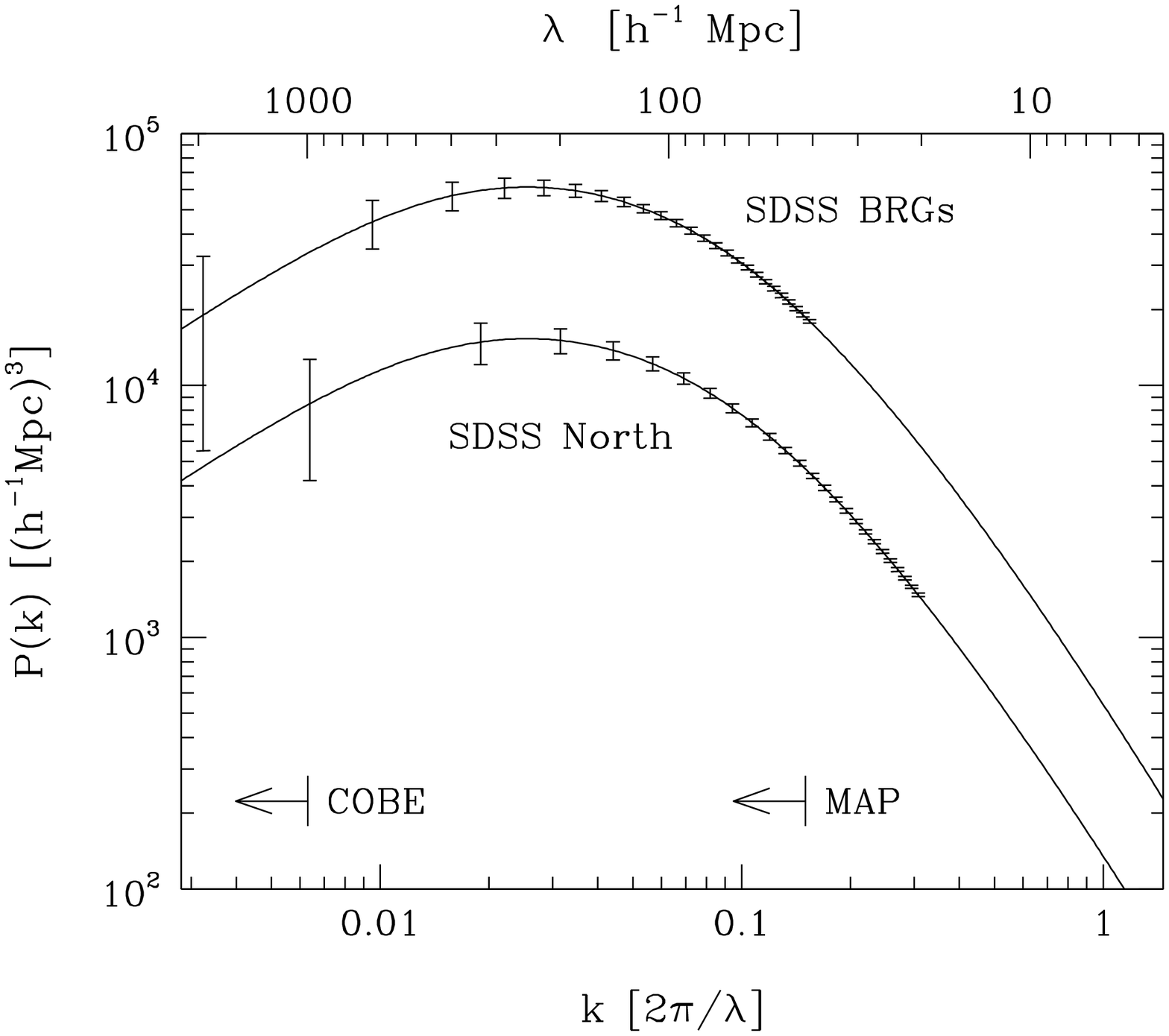}
\vspace{-5mm}
\caption[]{Left: (a) Expected $1\sigma$ uncertainty in the galaxy power spectrum
	measured from a volume-limited sample from the SDSS northern survey,
	along with predictions of $P(k)$ from four variants of the
	low-density CDM model.
	Note that the models have been arbitrarily normalised to agree
	on small scales ($k = 0.4$); in practice the COBE observations
	of CMB fluctuations fix the amplitude of $P(k)$ on very large scales.
	Right: (b) Power spectrum expected from the luminous red galaxy sample
	(BRGs), assuming that these galaxies have a bias factor twice that of 
	the flux-limited galaxy sample.}
\label{fig:P_k}
\end{figure}

The huge volume of the SDSS redshift survey will enable
estimates of the galaxy power spectrum to $\sim 1000 \hMpc$ scales.
Figure~\ref{fig:P_k}a shows the power spectrum $P(k)$ we would expect
to measure from a volume-limited ($M < M^*$) sample of galaxies from
the SDSS northern redshift survey, assuming Gaussian fluctuations
and a $\Omega h = 0.3$ CDM model.
The error bars include cosmic variance and shot noise, but not systematic
errors, due, for example, to galactic obscuration.
Provided such errors can be corrected for, (and star colours in the Sloan
survey will provide our best {\em a posteriori}
estimate of galactic obscuration),
then the figure shows that
we can easily distinguish between $\Omega h = 0.2$ and $\Omega h = 0.3$ models,
just using the northern main galaxy sample.
Adding the the luminous red galaxy sample
(Fig.~\ref{fig:P_k}b), will
further decrease measurement errors on the largest scales, and so we also
expect to be able to easily distinguish between low-density CDM and MDM models,
and models with differing indices $n$ for the shape of the primordial
fluctuation spectrum.

\subsection{Cosmological Density Parameter}

By measuring the distortions introduced by streaming motions
into redshift-space measures of galaxy clustering, one can constrain the
parameter $\beta = \Omega^{0.6}/b$, where $\Omega$ is the cosmological 
density paramter and $b$ is the bias factor relating fluctuations in
galaxy number density to fluctuations in the underlying mass distribution.
While existing redshift surveys, eg. IRAS \cite{cfw95} and Stromlo-APM
\cite{lemp96}, are hinting
that $\beta < 1$ (ie. that galaxies are significantly biased tracers
of mass or that $\Omega < 1$), their volumes are too small to
measure galaxy clustering in the fully linear regime reliably enough
to measure $\beta$ to much better than 50\% or so.
With the SDSS redshift survey, we expect to be able to constrain
$\beta$ to 10\% or better.

There are several ways we might hope to determine the galaxy bias factor $b$.
By measuring galaxy clustering on $\sim 1000 \hMpc$ scales as shown in
Figure~\ref{fig:P_k}, we can compare
with the COBE microwave background fluctuations directly, and so
constrain large-scale galaxy bias in a model-independent way.
Analysis of higher-order clustering statistics \cite{gf94},
and of non-linear dynamical effects \cite{cfw95} will also set constraints
on galaxy bias.
Knowing $\beta$ and $b$. we will be in a good position to reliably
measure the cosmological density parameter
$\Omega$ independent of models for the shape of the fluctuation spectrum.

\section{Conclusions}

It is probably no exaggeration to claim that the Sloan Digital Sky Survey
will revolutionize the field of large scale structure.
Certainly we can expect to rule out large numbers of presently viable 
cosmological models, as illustrated in Figure~\ref{fig:P_k}.
As well as measuring redshifts for a carefully controlled sample of
$10^6$ galaxies and $10^5$ quasars, the survey will also provide high quality
imaging data for about 100 times as many extragalactic objects,
from which one can obtain colour and morphological information.
In addition to measuring the basic cosmological parameters $\Omega$ and $h$
discussed in the preceding section, the SDSS will also allow us to measure the 
properties of galaxies as a function
of their colour, morphology and environment, providing valuable clues
to the process of galaxy formation.

The work described here has been carried out by people throughout
the SDSS collaboration.
It is a pleasure to thank the organisers of Dark98 for a most enjoyable
meeting and for supporting my attendance.

\end{document}